\documentclass[twocolumn,superscriptaddress]{revtex4}

\usepackage{amssymb}
\usepackage{amsmath}

\usepackage{graphicx}
\usepackage{dcolumn}
\usepackage{bm}

\newfont{\Fr}{eufm10}

\begin{document}

\title{Observational information for f(T) theories and Dark Torsion}
\author{Gabriel R. Bengochea}
\email{gabriel@iafe.uba.ar}
\affiliation{Instituto de Astronom\'\i a y F\'\i sica del Espacio
(IAFE), CC 67, Suc. 28, 1428 Buenos Aires, Argentina}

\begin{abstract}

In the present work we analyze and compare the information coming
from different observational data sets in the context of a sort of
f(T) theories. We perform a joint analysis with measurements of
the most recent type Ia supernovae (SNe Ia), Baryon Acoustic
Oscillation (BAO), Cosmic Microwave Background radiation (CMB),
Gamma-Ray Bursts data (GRBs) and Hubble parameter observations
(OHD) to constraint the only new parameter these theories have. It
is shown that when the new combined BAO/CMB parameter is used to
put constraints, the result is different from previous works. We
also show that when we include Observational Hubble Data (OHD) the
simpler LambdaCDM model is excluded to one sigma level, leading
the effective equation of state of these theories to be of phantom
type. Also, analyzing a tension criterion for SNe Ia and other
observational sets, we obtain more consistent and better suited
data sets to work with these theories.

\end{abstract}

\pacs{Valid PACS appear here}  \maketitle

\section{Introduction}

Current cosmological observations, mainly from type Ia supernovae,
show that the universe is undergoing accelerated expansion
\cite{perlmutter-otros, kowalski, union2, wmap7}. This accelerated
expansion has been attributed to a dark energy component with
negative pressure. The simplest explanation for this dark energy
seems to be the cosmological constant. However, among many
candidates \cite{sahni, padma, frieman}, some modified gravity
models have also been proposed based on, for example, $f(R)$
theories \cite{buch, staro, ker, barrow1, barrow2, carroll,
starobinsky, hu, odintsov1, odintsov2, capo, sotiriou}.

Some models based on modified teleparallel gravity were presented
as an alternative to inflationary models \cite{franco,franco2} or
showing a cosmological solution for the acceleration of the
universe by means of a sort of theories of modified gravity,
namely $f(L_T)$ \cite{tortion}, based on a modification of the
Teleparallel Equivalent of General Relativity (TEGR) Lagrangian
\cite{einstein, hayashi} where \emph{dark torsion} is the
responsible for the observed acceleration of the universe, and the
field equations are always 2nd order equations. It was shown in
\cite{tortion} that this fact makes these theories simpler than
the dynamical equations resulting in $f(R)$ theories among other
advantages. Recently, in \cite{linder} this sort of modified
gravity theories was called $f(T)$ theories and some works have
begun to develop in this area
\cite{wu,wu2,wu3,myr,yer,yang,karami,dent,lib,bamba}.

In \cite{nesseris} the tension and systematics in the Gold06 SNe
Ia data set have been investigated in great detail. Other authors,
working with different SNe Ia sets found these were in tension
with other SNe Ia sets and also with BAO and CMB \cite{wei1, li}.
In \cite{wei1}, analyzing the Union data set \cite{kowalski}, the
UnionT truncated data set was built by discarding the supernovae
generating the tension by using the $\Lambda$CDM model to select
the outliers. In \cite{li}, performing the same truncation
procedure of \cite{wei1} for 10 different models, it was suggested
that the impact of different models would be negligible.

In this work we present thorough observational information useful
to work with $f(T)$ theories by using the latest Union2 SNe Ia
compilation released \cite{union2}, the new combined parameter
from Baryon Acoustic Oscillation and Cosmic Microwave Background
radiation (BAO/CMB) \cite{sollerman} (more suitable for
non-standard models than the usually used $R$ and $A$ parameters),
a Gamma-Ray Burst data set \cite{xu2} and constraints from
Observational Hubble Data (OHD) \cite{ohd39, ohd40, ohd41}.

This Letter is organized as follows: in Section II we review the
fundamental concepts about $f(T)$ theories to, in Section III,
analyze a criterion of tension to improve the study of the new
data sets including BAO/CMB and GRBs. In Section IV we perform the
truncation of Union2 calculating the relative deviation to the
best fit of the $f(T)$ prediction for each one of the 557 points
following \cite{wei1, li} in order to show the disappearing of
tension and establishing a new set suitable for $f(T)$ theories.
In Section V we add the OHD observational information and discuss
some remarkable results and, in Section VI, we summarize the
conclusions of this work.

\section{General considerations about $f(T)$ theories}

Teleparallelism \cite{einstein, hayashi} uses as dynamical object
a vierbein field ${\mathbf{e}_i(x^\mu)}$, $i=0, 1, 2, 3$, which is
an orthonormal basis
for the tangent space at each point $x^\mu$ of the manifold: $\mathbf{e}%
_i\cdot\mathbf{e}_j=\eta_{i\, j}$, where $\eta_{i\, j}=diag
(1,-1,-1,-1)$.
Each vector $\mathbf{e}_i$ can be described by its components $e_i^\mu$, $%
\mu=0, 1, 2, 3$ in a coordinate basis; i.e.
$\mathbf{e}_i=e^\mu_i\partial_\mu $. Notice that Latin indices
refer to the tangent space, while Greek indices label coordinates
on the manifold. The metric tensor is obtained from the dual
vierbein as $g_{\mu\nu}(x)=\eta_{i\, j}\, e^i_\mu (x)\, e^j_\nu
(x)$. Differing from General Relativity (GR), which uses the
torsionless Levi-Civit\`{a} connection, Teleparallelism uses the
curvatureless Weitzenb\"{o}ck connection \cite{weit}, whose non-null
torsion is
\begin{equation}  \label{torsion2}
{T}^\lambda_{\:\mu\nu}=\hat{\Gamma}^\lambda_{\nu\mu}-\hat{\Gamma}^\lambda_{\mu\nu}=e^\lambda_i\:(\partial_\mu
e^i_\nu-\partial_\nu e^i_\mu)
\end{equation}
The TEGR Lagrangian is built with the torsion (\ref{torsion2}),
and its dynamical equations for the vierbein imply the Einstein
equations for the metric. The teleparallel Lagrangian is
\cite{hayashi,maluf,arcos},
\begin{equation}  \label{lagTele}
L_T\equiv T=S_\rho^{\:\:\:\mu\nu}\:T^\rho_{\:\:\:\mu\nu}
\end{equation}
where:
\begin{equation}  \label{S}
S_\rho^{\:\:\:\mu\nu}=\frac{1}{2}\Big(K^{\mu\nu}_{\:\:\:\:\rho}+\delta^\mu_%
\rho \:T^{\theta\nu}_{\:\:\:\:\theta}-\delta^\nu_\rho\:
T^{\theta\mu}_{\:\:\:\:\theta}\Big)
\end{equation}
and $K^{\mu\nu}_{\:\:\:\:\rho}$ is the contorsion tensor:
\begin{equation}  \label{K}
K^{\mu\nu}_{\:\:\:\:\rho}=-\frac{1}{2}\Big(T^{\mu\nu}_{\:\:\:\:\rho}
-T^{\nu\mu}_{\:\:\:\:\rho}-T_{\rho}^{\:\:\:\:\mu\nu}\Big)
\end{equation}
which equals the difference between Weitzenb\"{o}ck and Levi-Civit\`{a}
connections.

For a flat homogeneous and isotropic Friedmann-Robertson-Walker
universe (FRW),
\begin{equation}  \label{tetradasFRW}
e^i_\mu=diag(1,a(t),a(t),a(t))
\end{equation}
where $a(t)$ is the cosmological scale factor. By replacing in
(\ref{torsion2}), (\ref{S}) and (\ref{K}) one obtains
\begin{equation}  \label{STFRW}
T=S^{\rho\mu\nu}T_{\rho\mu\nu}=-6\:\frac{\dot{a}^2}{a^2}=-6\:H^2
\end{equation}
$H$ being the Hubble parameter $H=\dot{a}\, a^{-1}$.

In these modified gravity theories, the action is built promoting
$T$ to a function $f(T)$. The case $f(T)=T$ corresponds to TEGR.
In an $f(T)$ theory the spinless matter couples to the metric in
the standard form. Therefore, the equations of a freely falling
particle are the equations of the geodesics. Moreover, the source
in the equations for the geometry results to be the matter
energy-momentum tensor. In these aspects there is no difference
with GR. If matter is distributed isotropically and homogeneously,
the metric is the FRW metric and all kinematic equations
(luminosity distance, angular distance, cosmological redshift,
etc.) will be identical to the GR case. Any modification in the
null geodesics followed by light rays will be exclusively in the
scale factor $a(t)$. Some authors have mentioned that $f(T)$
theories are not invariant under local Lorentz transformations
\cite{franco,lib}. However, if this would affect the viability of
these models is a subject which is currently being analyzed.

The variation of the action with respect to the vierbein leads to
the field equations,
\begin{eqnarray}\nonumber
e^{-1}\partial_\mu(e\:S_i^{\:\:\:\mu\nu})f^{\prime}(T)-e_i^{\:\lambda}
\:T^\rho_{\:\:\:\mu\lambda}\:S_\rho^{\:\:\:\nu\mu}f^{\prime}(T)+\\
S_i^{\:\:\:\mu\nu}\partial_\mu(T)f^{\prime\prime}(T)+\frac{1}{4}
\:e_i^\nu \:f(T)=4\:\pi\:G\:e_i^{\:\:\:\rho}\:T_\rho^{\:\:\:\nu}
\label{ecsmovim}
\end{eqnarray}
where a prime denotes differentiation with respect to $T$,
$S_i^{\:\:\mu\nu}=e_i^{\:\:\rho}S_\rho^{\:\:\mu\nu}$ and
$T_{\mu\nu}$ is the matter energy-momentum tensor.

The substitution of the vierbein (\ref{tetradasFRW}) in (\ref%
{ecsmovim}) for $i=0=\nu$ yields
\begin{equation}  \label{FRWMod}
12\:H^2\:f^{\prime}(T)+f(T)=16\pi G\:\rho
\end{equation}
Besides, the equation $i=1=\nu$ is
\begin{equation}  \label{ec1-1}
48 H^2 f^{\prime\prime}(T) \dot{H}-f^{\prime}(T)[12 H^2 + 4 \dot{H}%
]-f(T)=16\pi G\:p
\end{equation}
In Eqs. (\ref{FRWMod}-\ref{ec1-1}), $\rho(t)$ and $p(t)$ are the
total density and pressure respectively.

In \cite{tortion} it was shown that when $f(T)$ is a power law
such as
\begin{equation}
f(T)=T-\frac{\alpha }{(-T)^{n}}  \label{fT}
\end{equation}%
leads to reproduce the observed accelerated expansion of the
universe, being $\alpha $ and $n$ real constants to be determined
by observational constraints.

From (\ref{FRWMod}) along with (\ref{fT}), the modified Friedmann
equation results to be (e.g. \cite{tortion})
\begin{equation}
H^{2}-\frac{(2n+1)\;\alpha }{6^{n+1}H^{2n}}=\frac{8}{3}\pi G\rho
\label{friedmodif}
\end{equation}%
where $\rho =\rho _{mo}(1+z)^{3}+\rho _{ro}(1+z)^{4}$, $z$ is the
cosmological redshift and as it is usual, we will call $\Omega
_{i}=8\pi G\,\rho _{io}/(3H_{0}^{2})$ to the contributions of
matter and radiation to the total energy density today. For
$\alpha =0$ the GR spatially flat Friedmann equation is retrieved.
The case $n=0$ recovers the GR dynamics with cosmological
constant. Compared with GR, $n$ is the sole new free parameter
(see \cite{tortion} for details).

In the next sections, we will use a
$\chi^2=\chi^2_{SNe}+\chi^2_{BAO/CMB}+\chi^2_{GRB}+\chi^2_{OHD}$
statistic to find best fits for the free parameters $\Omega_m$ and
$n$ of a model given by (\ref{fT}) using several data sets. The
separate $\chi^2$ of SNe Ia, BAO/CMB, GRBs and OHD and the
corresponding data sets used in this work are shown in Appendix B.
In order to see whether our model is favored over the $\Lambda$CDM
model, we will also use the information criterion known as $AIC$
(Akaike Information Criterion) \cite{akaike, liddle}. The $AIC$ is
defined as $AIC=-2\:ln {\cal L}_{max}+2k$, where the likelihood is
defined as ${\cal L}\propto e^{\chi^2/2}$, the term $-2\:ln {\cal
L}_{max}$ corresponds to the $\chi^2_{min}$ and $k$ is the number
of parameters of the model. According to this criterion a model
with the smaller $AIC$ is considered to be the best, and a
difference $|\Delta AIC|$ in the range between 0 and 2 means that
the two models have about the same support from the data. For a
difference between 2 and 4 this support is considerably less for
the model with the larger $AIC$, while for a difference $>$10 the
model with the larger $AIC$ is practically irrelevant \cite{bies}.

\section{Constraining Dark Torsion with updated data sets}

We found interesting to analyze what would happen if we applied a
criterion in order to study the consistency between data sets, a
criterion more restrictive than the only fact that the confidence
intervals overlap. To perform this analysis, we adopted the
criterion of considering the existence of tension between a given
data set and another set constituted combining several data sets
(including the first one) as the fact that the best fit point to
the first data set is out of the 68.3\% (1$\sigma$) confidence
level contour given by the combined data set. Similar criteria
were adopted in their analysis by \cite{nesseris, wei1, li}. One
could choose not to use this more restrictive criterion; however,
we wanted to investigate its consequences of applying it to
several data sets in the framework of $f(T)$ theories. In
\cite{nesseris}, for example, the best fits to sets and subsets of
SNe are compared with the means of determining if two of those are
in tension or not, and how far from the confidence intervals lies
the $\Lambda$CDM model. With our adopted criterion, we seek more
physical consistency between best-fits, so the best fits do not
drive to too different cosmological evolutions. The best fit which
effective equation of state is of the phantom type \cite{cal02}
($w_{eff}<-1$) tells us about very different physics from the one
that is not. Also, best fits that lie too far apart from
$\Lambda$CDM model ($n=0$ or $w_{eff}$=-1) will indicate the need
of more exotic models.

Something important to consider is that the best fits to the SNe
or their combination depend also on the fitter used to process the
SNe data sets. Avoiding this type of tension we make sure that in
most cases both best fits (SNe Ia and combined data sets) have
similar results in the equation of state $w_{eff}$ or the $n$
parameter. In the $w$CDM ($w$=const) framework, for example, it
has been shown that with the SDSSII (MLCS) data set \cite{kessler,
mlcs} both best fits suggest different cosmic evolutions while
when relieving the tension this problem disappears \cite{gb2}.

\begin{figure}[h!]
\begin{center}
\includegraphics[width=7cm,angle=0]{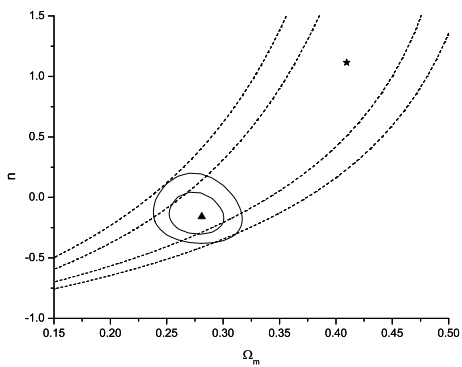}
\end{center}
\caption{Confidence intervals at 68.3\% and 95.4\% in the
$\Omega_m-n$ plane for the UnionT observations of SNe Ia only
(dashed lines) and UnionT SNe Ia, BAO and CMB (solid lines). We
also show the best fit values to the observations of UnionT SNe Ia
only (star), and to the combination of UnionT SNe Ia, BAO and CMB
(triangle).} \label{fig1}
\end{figure}

Then, with this criterion, in our previous work \cite{tortion}
existed some tension between the data from the SNe Ia Union sample
\cite{kowalski} and BAO \cite{eisen} and CMB \cite{komatsu}.
Taking the truncated UnionT data set from \cite{wei1} we proceeded
to evaluate the possible tension between UnionT, BAO and CMB.

We found there was still tension between these data sets as shown
in Fig. \ref{fig1}. On one hand, the best fit to SNe Ia is of the
phantom type and the one corresponding to the combined set is not
and, on the other hand, the best fit to SNe Ia drives $H$ in
recombination to be 20\% greater than for the best fit to
SNe+BAO+CMB, and $w_{eff}$ today to be 35\% greater. Considering
the mentioned criterion, UnionT eliminates the tension between
data in some models as shown in \cite{li}, but it is not the case
with $f(T)$.

Instead of continuing truncating the Union set, now we proceeded
to use the latest SNe Ia data set Union2 \cite{union2} (processed
with SALT2 light-curve fitter \cite{salt2}) and the combined
parameter BAO/CMB. This combined BAO/CMB parameter implemented in
\cite{sollerman} is more suitable to add constraints to
non-standard models (see Appendix B for details).

The best fit to the Union2 SNe Ia data set only, was achieved with
$n=0.49$ and $\Omega_m=0.33$ with the reduced
$\chi^2_{min}/\nu\simeq0.98$ (or equivalently,
$\Delta\chi^2_{min}=-0.18$ with respect to $\Lambda$CDM with
$\Omega_m=0.27$ \cite{wmap7}), where $\nu$ is the number of
degrees of freedom. All the results with their corresponding
1$\sigma$ uncertainties and the analysis from the $AIC$ criterion
are summarized in Table \ref{results}.

Working only with the new BAO/CMB parameter we found the value of
$n$ for the best fit is remarkably higher than with other data
sets ($n=4.58$), although more efficient in constraining
$\Omega_m$ (in contrast to working with BAO and CMB parameters
separately) having a range of values more consistent with other
observations, such as weak lensing and its combination with CMB
and SNe Ia (e.g. \cite{lensing1, lensing2}). In our case we found
$\Omega_m=0.28$. These values ($n=4.58$, $\Omega_m=0.28$) perform
a better fit than $\Lambda$CDM by a $\Delta\chi^2_{min}=-1.46$.
Comparing these results with our previous work \cite{tortion}, we
also found that including this combined parameter, the value of
$n$ for the best fit using BAO/CMB or its combination with the
rest of the data sets results always greater than zero. Therefore,
the effective dark torsion is of the phantom type \cite{cal02}.
This result is also in opposition with recent constraints when BAO
and CMB are used separately through the parameters $A$ and $R$
respectively \cite{wu}. Combining the SNe Ia data with BAO/CMB
data we found the best fit, which can be seen in Table
\ref{results}.

We added to our analysis a data set of observations of Gamma-Ray
Bursts (GRBs). Knowing that there are still debates about if these
objects are standard candles (e.g. \cite{ghi, basi}), we followed
the policy assumed in other published works such as \cite{wei2,
xu2} to see how our results are modified and to check the
consistency of both approaches presented by \cite{wei2} and
\cite{xu2} in the framework of $f(T)$ theories. Recently, in
\cite{xu1} it has been demonstrated that the data set of
\cite{xu2} is consistent with Union2. One could do the analysis
using data sets consisting of SNe Ia and GRBs separately.
Otherwise, an analysis can be made compiling Union2 and one GRBs
data sets together \cite{wei2}. These two separate analysis showed
identical results (see Appendix A).

Here we used the approach developed in \cite{xu2} firstly for
being a set-independent data set, this means this set is not only
applicable with Union2, but with any other SNe Ia data set. Most
importantly, we used this data set because evaluating it
separately from the SNe Ia is more helpful in our work of finding
tension between SNe Ia data and other data sets. Otherwise, in
case of finding tension we might have needed to truncate a
combination of SNe+GRBs data set, being this last a sum of
different data. When incorporating GRBs to the joint statistic we
observed that the addition of the mentioned observations slightly
reduce the size of the confidence intervals and the joint best fit
is displayed in Table \ref{results}.

\begin{figure}[h!]
\begin{center}
\includegraphics[width=7cm,angle=0]{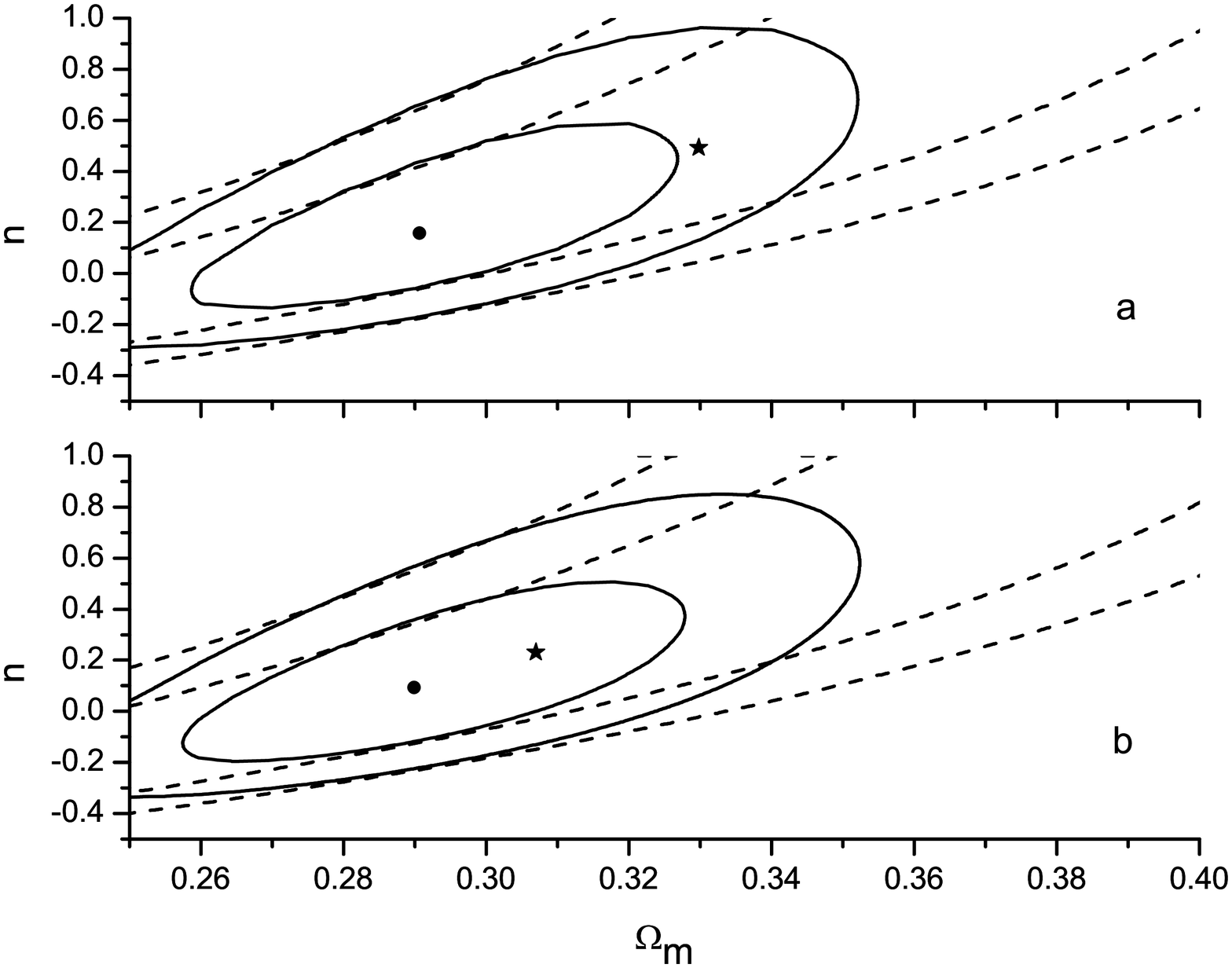}
\end{center}
\caption{(a) Confidence intervals at 68.3\% and 95.4\% in the
$\Omega_m-n$ plane for the Union2 observations of SNe Ia only
(dashed lines) and Union2 SNe+BAO/CMB+GRBs (solid lines). We also
show the best fit values to the observations of Union2 SNe Ia only
(star), and to the combination of Union2 SNe+BAO/CMB+GRBs (dot).
(b) Idem (a), for the Union2T data set. } \label{snvssuma}
\end{figure}

In Fig. \ref{snvssuma}a, we show the confidence intervals of SNe
Ia with the combination of SNe+BAO/CMB+GRBs. There, it can be
easily seen that with the adopted criterion for the tension
between data sets, there is a slight tension between Union2 and
the other data sets.

In the next section, we will adopt the criterion presented in
\cite{wei1} in order to perform the truncation of Union2 SNe Ia
data set with the objective of dissipating the tension between
data sets when analyzing these $f(T)$ type of theories.

\section{Building a new improved data set}

We followed the simple method used in \cite{nesseris, wei1} to
find the outliers responsible for the tension. In \cite{nesseris}
the distance moduli of the six SNe Ia which are mostly responsible
for the tension in Gold06 data set differ by more than $1.8
\sigma$ from the $\Lambda$CDM prediction. In \cite{wei1} 21 SNe Ia
were discarded to build the UnionT set in order to eliminate the
tension with CMB and BAO.

Similarly, we firstly fitted our $f(T)$ model to the whole 557 SNe
Ia in the Union2 data set and found the best fit parameters (with
the corresponding $\mu_0=43.15$). Then, we calculated the relative
deviation to the best fit prediction,
$\mid\mu_{obs}-\mu_{th}\mid/\sigma_{obs}$, for all the 557 data
points. We found that as in \cite{wei1} the cut $1.9\sigma$ solved
the tension problem. This cut implied to take out 39 SNe Ia from
Union2. With the remaining 518 SNe Ia we build the Union2T data
set. The outliers are shown in Table \ref{outliers}.
\begin{table}[h!]
\caption{The names of the outliers from Union2 data set.}

\centering
\begin{tabular}{|c|} \hline\hline
\bf Outliers from Union2 \\
\hline 1995ac, 1998dx, 1999bm, 2001v, 2002bf, 2002hd, 2002hu\\
2002jy, 2003ch, 2003ic, 2006br, 2006cm, 2006cz, 2007ca\\
10106, 2005ll, 2005lp, 2005fp, 2005gs, 2005gr, 2005hv\\
2005ig, 2005iu, 2005jj, 1997k, 1998ba, 03D4au, 04D3cp\\
04D3oe, 03D4cx, 03D1co, d084, e140, f308, g050\\
g120, m138, 05Str, 2002fx \\
\hline\hline
\end{tabular}
\label{outliers}
\end{table}

In Fig. \ref{snvssuma}b, it is shown the result of using Union2T
with the combination of Union2T+BAO/CMB+GRBs. The best fit to
Union2T was achieved with the values $n=0.23$, $\Omega_m=0.31$ and
with $\chi^2_{min}/\nu=0.67$ ($\Delta\chi^2_{min}=-0.20$), whilst
the best fit to the joint analysis of Union2T+BAO/CMB+GRBs was
obtained with $n=0.09$, $\Omega_m=0.29$ with
$\chi^2_{min}/\nu=0.67$ ($\Delta\chi^2_{min}=-0.97$). From these
results we can see that the $\chi^2_{min}$ as the
$\chi^2_{min}/\nu$ have been significantly improved in respect to
the values obtained in the previous section.

This Union2T set along with the corresponding sets of BAO/CMB and
GRBs are more consistent between them, considering the adopted
criterion which advantages were mentioned above.

\section{Adding OHD data}

In this section we wondered about how the results of the previous
sections were modified when a data set with Hubble parameter
observations $H(z)$ was added to the $\chi^2$ statistic. The
details of the used data are displayed in Appendix B.

\begin{figure}[h!]
\begin{center}
\includegraphics[width=7cm,angle=0]{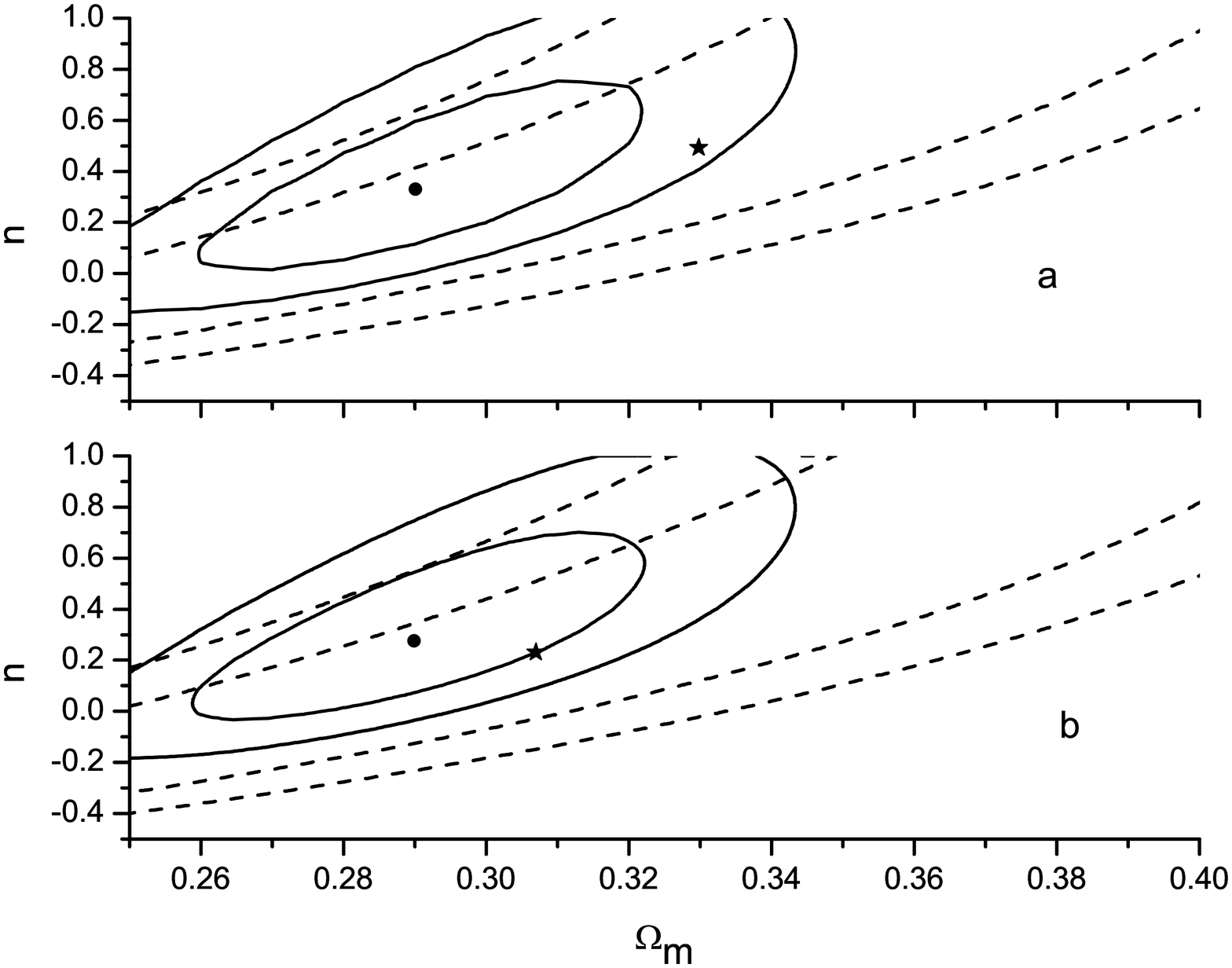}
\end{center}
\caption{(a) Confidence intervals at 68.3\% and 95.4\% in the
$\Omega_m-n$ plane for the Union2 observations of SNe Ia only
(dashed lines) and Union2 SNe+BAO/CMB+GRBs+OHD (solid lines). We
also show the best fit values to the observations of Union2 SNe Ia
only (star), and to the combination of Union2 SNe+BAO/CMB+GRBs+OHD
(dot). (b) Idem (a), for the Union2T observations.}
\label{u2vstotal}
\end{figure}

In Fig. \ref{u2vstotal}a, the confidence intervals are shown at
68.3\% and 95.4\% for the SNe Ia data only and for the combination
of SNe+BAO/CMB+GRBs+OHD. Also, displaying the corresponding best
fits. This analysis showed that the best fit was reached with
$n=0.33$ and $\Omega_m=0.29$ with a $\chi^2_{min}/\nu=0.96$
($\Delta\chi^2_{min}=-2.6$). It is observed that the confidence
intervals are slightly smaller in size than when OHD data is not
added. Surprisingly, we found that the adding of OHD pushes the
$\Lambda$CDM model out of the $1\sigma$ confidence level of the
combined data. The inclusion of OHD data, then, favors an equation
of state of the phantom type and makes the values that lie at
68.3\% to be $n \in [0.12, 0.58]$, $\Omega_m \in [0.27, 0.31]$. A
similar result was obtained when combining BAO/CMB, the shift
parameter $R$ of CMB and supernovae Ia from SDSS by using the MLCS
light-curve fitter (Fig. 1 of \cite{sollerman}).

Again, with the adopted criterion about the existence of tension
between data sets we see there is a slight tension between Union2
and Union2 SNe+BAO/CMB+GRBs+OHD. Performing the analogue procedure
of the previous section, we found that a $1.9\sigma$ cut was
suitable to remove the tension between data sets. The result is
displayed in Fig. \ref{u2vstotal}b.

The combination Union2T SNe Ia with BAO/CMB, GRBs and OHD allowed
the $\Lambda$CDM model again to lie inside the $1\sigma$ joint
probability region.

We summarize in Table \ref{results} the main results from the
analysis performed in this work.

\begin{table}[h!]
\caption{Best fit values and 1$\sigma$ errors for each parameter
marginalizing over the other, for the models considered in this
work. (1) SNe, (2) BAO/CMB, (3) SNe + BAO/CMB, (4) SNe + BAO/CMB +
GRBs, (5) SNe + BAO/CMB + GRBs + OHD, (6) U2T, (7) U2T + BAO/CMB +
GRBs and (8) U2T + BAO/CMB + GRBs + OHD. SNe stands for SNe Ia
from Union2 data set, U2T stands for the truncated Union2 data set
with a $1.9\sigma$ cut and $\Delta AIC=AIC_{f(T)}-AIC_{\Lambda
CDM}$.}

\centering
\begin{tabular}{|c|c|c|c|c|c|} \hline\hline
\bf Data set(s) &\bf $n$ & \bf $\Omega_m$ &\bf $\chi^2_{min}/\nu$ & $\Delta\chi^2_{min}$ & $\Delta AIC$\\
\hline\hline 1 & $0.49^{+1.13}_{-1.09}$  & $0.33^{+0.08}_{-0.19}$ & 0.98 & -0.18 & 1.82\\
2 & $4.58^{+n/a}_{-4.87}$ & $0.28^{+0.02}_{-0.02}$ & - & -1.46 & 0.54\\
3 & $0.15^{+0.28}_{-0.18}$ & $0.29^{+0.02}_{-0.02}$ & 0.97 & -0.99 & 1.00\\
4 & $0.16^{+0.25}_{-0.18}$ & $0.29^{+0.02}_{-0.02}$ & 0.96 & -0.93 & 1.07\\
5 & $0.33^{+0.25}_{-0.21}$ & $0.29^{+0.02}_{-0.02}$ & 0.96 & -2.6 & -0.60\\
6 & $0.23^{+1.08}_{-0.49}$ & $0.31^{+0.08}_{-0.18}$ & 0.67 & -0.20 & 1.80\\
7 & $0.09^{+0.25}_{-0.18}$ & $0.29^{+0.02}_{-0.02}$ & 0.67 & -0.97 & 1.03\\
8 & $0.28^{+0.24}_{-0.21}$ & $0.29^{+0.02}_{-0.02}$ & 0.67 & -1.83 & 0.17\\
\hline\hline
\end{tabular}
\label{results}
\end{table}

\section{Conclusions}

We have updated the constraints to an $f(T)=T-\alpha(-T)^{-n}$
theory by using the latest type Ia supernovae data set Union2, the
new combined BAO/CMB parameter, a Gamma-Ray Bursts set and Hubble
Observational Data.

When the new BAO/CMB parameter is used instead of the frequently
used $A$ and $R$ parameters separately the best fit values change
with respect to previous works. From Table \ref{results} we see
that all best fit values are for $n>0$, leading the effective
equation of state to be phantom like. Note that in all cases where
SNe Ia data were involved, we used the Union2 data set which was
processed with SALT2 fitter and this could be an additional factor
in the obtained results as showed in \cite{gb2}. Adding GRB data
did not modify the results appreciatively and we also found that
two approximations performed by different authors are consistent
between them and lead to the same results.

We found that when including information from OHD to put
constraints, as in the BAO/CMB case, an equation of state of the
phantom type is favored. Remarkably, the simpler $\Lambda$CDM
model lies outside the 68.3\% confidence level region of the
combined SNe+BAO/CMB+GRBs+OHD data. The values that lie at 68.3\%
are in the ranges $n \in [0.12, 0.58]$, $\Omega_m \in [0.27,
0.31]$.

The adopted criterion of tension between data sets in this work
and the truncation process performed to Union2 data set allowed
us, firstly, that the physics that determines the cosmological
evolution through the $n$ parameter does not differ much when only
the SNe are considered or when those are combined with other data
sets, and secondly, that each data set is consistent amongst the
others. In every case, eliminating tension leaded to reduce at
least to one half the ratio between values of $w_{eff}$ in $z=0$
and $H$ in recombination, obtained from the best fits to SNe alone
or their combination with other data sets. Additionally, the
removal of tension by this criterion resulted in, when combining
Union2T data set with BAO/CMB+GRB+OHD, the $\Lambda$CDM model to
lie inside the 1$\sigma$ joint probability region of the data set
with all observations. From Table \ref{results}, the values for
$\Delta\chi^2_{min}$ always favored the $f(T)$ models, whilst when
performing an analysis with the $AIC$ criterion we found that all
the evaluated cases of $f(T)$ have the same support from the data
with respect to the $\Lambda$CDM model since $0<|\Delta AIC|<2$.

\bigskip

\acknowledgments{G.R.B. is supported by CONICET. I would like to
thank Eric Linder for kind dialogues, to Ariel Goobar, Hao Wei and
Lixin Xu for helpful discussions. Also, I thank Diego Travieso for
his numerical collaboration and interesting discussions and Rafael
Ferraro for his support.}

\appendix

\section{Analysis from Hymnium Gamma-Ray Bursts data set}

Here we show the result of adding to the performed analysis with
the Union2 SNe Ia data set, the set Hymnium of 59 GRBs according
to \cite{wei2}.

\begin{figure}[h!]
\begin{center}
\includegraphics[width=6cm,angle=0]{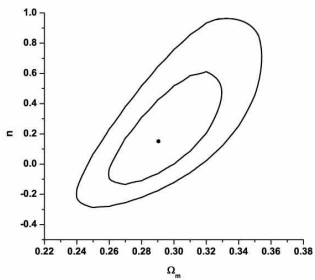}
\end{center}
\caption{Confidence intervals at 68.3\% and 95.4\% in the
$\Omega_m-n$ plane coming from combining SNe+GRBs+BAO/CMB data,
where the 59 GRBs from the Hymnium data set were added to Union2
according to \cite{wei2}.} \label{grbswei}
\end{figure}

With the objective of comparing the obtained results when adding
GRBs to SNe Ia data and BAO/CMB using the 5 values set of
\cite{xu2} and when adding the data sets in the way performed in
\cite{wei2}, we observed the differences in the 68.3\% and 95.4\%
confidence intervals of both methods. The result using the 59 GRBs
set of \cite{wei2} is displayed in Fig. \ref{grbswei}. The values
in the joint best fit are for $n=0.15^{+0.26}_{-0.18}$ and
$\Omega_m=0.29^{+0.02}_{-0.02}$. As can be seen, the result is
very similar to the one obtained in section III (Fig.
\ref{snvssuma}a).

\section{Cosmological constraints methods}

\subsection{Type Ia Supernovae constraints}

The data points of the 557 Union2 SNe Ia compiled in \cite{union2}
are given in terms of the distance modulus $\mu_{obs}(z_i)$ and
the corresponding uncertainty for each observed value
$\sigma(z_i)$. On the other hand, the theoretical distance modulus
is defined as
\begin{equation}\label{muth}
\mu_{th}(z_i)=5 log_{10} D_L(z_i)+\mu_0
\end{equation}
where $\mu_0\equiv 42.38 - 5 log_{10}h$ and $h$ is the Hubble
constant $H_0$ in units of 100 km/s/Mpc, whereas the Hubble-free
luminosity distance for the flat case is
\begin{equation}\label{dlHfree}
D_L(z)=
(1+z)\:\int_{0}^{z}\frac{dz^{\prime}}{E(z^{\prime},\mathbf{p})}
\end{equation}
in which $E\equiv H/H_0$, and $\mathbf{p}$ denotes the model
parameters (here, $n$ and $\Omega_m$). The parameter $\mu_0$ is a
nuisance parameter but it is independent of the data points.
Following \cite{chi2sn}, the minimization with respect to $\mu_0$
can be made by expanding the $\chi^2_{SNe}$ with respect to
$\mu_0$ as
\begin{equation}  \label{chib}
\chi^2_{SNe}(\mathbf{p})=\tilde{A}-2\mu_0 \tilde{B}+\mu_0^2
\tilde{C}
\end{equation}
where,
\begin{eqnarray}  \nonumber
\tilde{A}(\mathbf{p})&=& \sum_{i=1}^{N=557}\frac{[\mu_{obs}(z_i)-\mu_{th}(z_i;\mu_0=0,\mathbf{p})]^2}{\sigma^2(z_i)}\\
\tilde{B}(\mathbf{p})&=& \sum_{i=1}^{N=557}\frac{[\mu_{obs}(z_i)-\mu_{th}(z_i;\mu_0=0,\mathbf{p})]}{\sigma^2(z_i)}\nonumber\\
\tilde{C}&=& \sum_{i=1}^{N=557}\frac{1}{\sigma^2(z_i)} \nonumber
\end{eqnarray}
Equation (\ref{chib}) has a minimum for
$\mu_0=\tilde{B}/\tilde{C}$ at
\begin{equation}  \label{chi2posta}
\tilde{\chi}^2_{SNe}(\mathbf{p})=\tilde{A}(\mathbf{p})-\frac{\tilde{B}(\mathbf{p})^2}{\tilde{C}}
\end{equation}
Since $\chi^2_{SNe,min}=\tilde{\chi}^2_{SNe,min}$ obviously, we
can instead minimize $\tilde{\chi}^2_{SNe}$ which is independent
of $\mu_0$.

\subsection{Combined BAO/CMB parameter constraints}

In the $f(T)$ theories considered here (\ref{fT}), for later times
the term $-\alpha /(-T)^{n}$ is dominant, while in early times
when $H\rightarrow \infty $ General Relativity is recovered. Also,
since this model presents matter domination at the decoupling time
as the standard model, we can use the BAO and CMB information as
showed in \cite{tortion}.

When analyzing CMB and BAO observations there are two parameters
commonly employed, $R$ \cite{paramR} and $A$ \cite{eisen}.
However, a more model-independent constraint can be achieved by
multiplying the BAO measurement of $r_s(z_d)/D_V(z)$ with the
position of the first CMB power spectrum peak \cite{komatsu}
$\ell_A=\pi d_A(z_*)/r_s(z_*)$, thus cancelling some of the
dependence on the sound horizon scale \cite{sollerman}. Here,
$d_A(z_*)$ is the comoving angular-diameter distance to
recombination, $r_s$ is the comoving sound horizon at photon
decoupling, $z_d\thickapprox1020$ is the redshift of the drag
epoch at which the acoustic oscillations are frozen in, and $D_V$
is defined as (assumed a $\Lambda $CDM model) \cite{eisen},
\begin{equation}  \label{DV}
D_V(z)=\Bigg[\frac{z}{H(z)}\:\Big(\int_{0}^{z}\frac{dz^{\prime}}{%
H(z^{\prime})}\Big)^{2}\Bigg]^{1/3}
\end{equation}
We further assume $z_*=1090$ from \cite{komatsu} (variations
within the uncertainties about this value do not give significant
differences in the results).

In \cite{percival} was measured $r_s(z_d)/D_V(z)$ at two
redshifts, $z=0.2$ and $z=0.35$, finding
$r_s(z_d)/D_V(0.2)=0.1905\pm0.0061$ and
$r_s(z_d)/D_V(0.35)=0.1097\pm0.0036$. Combining this with $\ell_A$
gives the combined BAO/CMB constraints \cite{sollerman}:
\begin{eqnarray}\nonumber
\frac{d_A(z_*)}{D_V(0.2)}\:\frac{r_s(z_d)}{r_s(z_*)}&=& 18.32\pm0.59\\
\frac{d_A(z_*)}{D_V(0.35)}\:\frac{r_s(z_d)}{r_s(z_*)}&=&
10.55\pm0.35\label{soll}
\end{eqnarray}
Before matching to cosmological models we also need to implement
the correction for the difference between the sound horizon at the
end of the drag epoch and the sound horizon at last scattering.
The first is relevant for the BAO, the second for the CMB, and
$r_s(z_d)/r_s(z_*)=1.044\pm0.019$ (using values from
\cite{komatsu}). Inserting this into (\ref{soll}) and taking into
account the correlation between these measurements using the
correlation coefficient of 0.337 calculated by \cite{percival},
gives the final constraints we use for the cosmology analysis
\cite{sollerman}:
\begin{eqnarray}  \nonumber
A_1=\frac{d_A(z_*)}{D_V(0.2)}&=& 17.55\pm0.65\\
A_2=\frac{d_A(z_*)}{D_V(0.35)}&=& 10.10\pm0.38\label{bao/cmb}
\end{eqnarray}

Using this BAO/CMB parameter cancels out some of the dependence on
the sound horizon size at last scattering. This thereby removes
the dependence on much of the complex pre-recombination physics
that is needed to determine that horizon scale \cite{sollerman}.
In all the cases, we have considered a radiation component
$\Omega_r=5 x 10^{-5}$.

So, for our analysis we add to the $\chi^2$ statistic:
\begin{equation}  \label{chibaocmb}
\chi^2_{BAO/CMB}(\mathbf{p})=\sum_{i=1}^{N=2}\frac{[A_{obs}(z_i)-A_{th}(z_i;\mathbf{p})]^2}{\sigma_A^2(z_i)}
\end{equation}
where $\mathbf{p}=(n, \Omega_m)$ are the free parameters,
$A_{obs}$ is the observed value ($A_1$ and $A_2$), $A_{th}$ is the
predicted value by the model and $\sigma_A$ is the $1\sigma$ error
of each measurement.

\subsection{Gamma-Ray Bursts constraints}

Following \cite{schaefer} and \cite{xu2}, we consider the
well-known Amati's $E_{p,i}-E_{iso}$ correlation \cite{amati1,
amati2, amati3, amati4} in GRBs, where $E_{p,i}=E_{p,obs}(1+z)$ is
the cosmological rest-frame spectral peak energy, and $E_{iso}$ is
the isotropic energy.

In \cite{wang}, it was defined a set of model-independent distance
measurements $\{\bar{r}_p(z_i)\}$:
\begin{equation}  \label{rp1}
\bar{r}_p(z_i)\equiv\frac{r_p(z)}{r_p(z_0)}
\end{equation}
with,
\begin{equation}  \label{rp2}
r_p(z)\equiv \frac{(1+z)^{1/2}}{z}\:\frac{H_0}{c}\:r(z)
\end{equation}
where $r(z)=d_L(z)/(1+z)$ is the comoving distance at redshift
$z$, and $z_0=0.0331$ is the lowest GRBs redshift.

Following the method proposed by \cite{wang}, in \cite{xu2} were
obtained 5 model-independent distances data points and their
covariance matrix by using 109 GRBs via Amati's correlation. The
resulted model-independent distances $\bar{r}_p^{data}(z_i)$ and
their uncertainties, the correlation matrix and the covariance
matrix are these ones from \cite{xu2}.

So, a given cosmological model with $\{\mathbf{p}\}$ free
parameters can be constrained by GRBs via
\begin{eqnarray}  \label{chiGRB}
\chi_{GRB}^2(\mathbf{p})&=&[\Delta
\bar{r}_p(z_i)]^T\cdot(\mathbf{C}_{GRB}^{-1})\cdot[\Delta
\bar{r}_p(z_i)]\\
\Delta \bar{r}_p(z_i)&=&\bar{r}_p^{data}(z_i)-\bar{r}_p(z_i)
\label{deltarp}
\end{eqnarray}
where $\bar{r}_p(z_i)$ is defined by (\ref{rp1}) and
$\mathbf{C}_{GRB}^{-1}$ is the inverse of the covariance matrix.
In this way, the constraints for a large amount of observational
GRBs data is projected into the relative few quantities
$\bar{r}_p^{data}(z_i)$.

\subsection{Observational Hubble Data (OHD) constraints}

The Observational Hubble Data are based on differential ages of
the galaxies \cite{ohd65}. In \cite{ohd66} it was obtained an
independent estimate for the Hubble parameter using the method
developed in \cite{ohd65}, and the authors used it to constrain
the equation of state of dark energy. The Hubble parameter
depending on the differential ages as a function of redshift can
be written as,
\begin{equation}  \label{hohd}
H(z)=-\frac{1}{1+z}\:\frac{dz}{dt}
\end{equation}

So, once $dz/dt$ is known, $H(z)$ is obtained directly
\cite{ohd67}. By using the differential ages of passively-evolving
galaxies, in \cite{ohd67} was obtained $H(z)$ in the range of
$0.1\lesssim z \lesssim 1.8$ and in \cite{ohd41} new data at
$0.35<z<1$ were studied. Here, $H_0$ from \cite{ohd40} and eleven
observational Hubble data from \cite{ohd41} are used.

In addition, in \cite{ohd39} the authors took the BAO scale as a
standard ruler in the radial direction, called ''Peak Method'',
obtaining three more additional data (in km/s/Mpc):
$H(z=0.24)=79.69\pm2.32$, $H(z=0.34)=83.8\pm2.96$ and
$H(z=0.43)=86.45\pm3.27$, which are model- and scale-independent.
We just consider the statistical errors.

The best fit values of the cosmological model parameters from
observational Hubble data are then determined by minimizing,
\begin{equation}  \label{chiohd}
\chi^2_{OHD}(\mathbf{p})=\sum_{i=1}^{N=15}\frac{[H_{obs}(z_i)-H_{th}(z_i;\mathbf{p})]^2}{\sigma^2(z_i)}
\end{equation}
where as before, in this work $\mathbf{p}=(n, \Omega_m)$, $H_{th}$
is the predicted value for the Hubble parameter, $H_{obs}$ is the
observed value, $\sigma(z_i)$ is the standard deviation of each
measurement, and the summation is over the full 15 values at
redshift $z_i$ mentioned above.


\begin{thebibliography}{99}

\bibitem{perlmutter-otros} Perlmutter S. et al., Bull. Am. Astron. Soc.
\textbf{29}, (1997) 1351; Astrophys. J. \textbf{517}, (1999) 565;
Riess A. G. et al., Astron. J. \textbf{116}, (1998) 1009; Astron.
J. \textbf{607} (2004) 665.

\bibitem{kowalski}Kowalski M. et al., Astrophys. J. \textbf{686}, (2008) 749.

\bibitem{union2}Amanullah R. et al., Astrophys. J. \textbf{716}, (2010) 712.

\bibitem{wmap7}Komatsu E. et al., arXiv:1001.4538.

\bibitem{sahni} Sahni V. and Starobinsky A. A., Int. J. Mod. Phys. \textbf{D9}, (2000) 373.

\bibitem{padma} Padmanabhan T., Phys. Rep. \textbf{380}, (2003) 235.

\bibitem{frieman} Frieman J., Turner M. S. and Huterer D., Annu.
Rev. Astron. Astrophys. \textbf{46}, (2008) 385.

\bibitem{buch} Buchdahl H. A., Mon. Not. R. Astron. Soc. \textbf{150},
(1970) 1.

\bibitem{staro} Starobinsky A. A., Phys. Lett. \textbf{B91}, (1980) 99.

\bibitem{ker} Kerner R., Gen. Relat. Gravit. \textbf{14}, (1982) 453.

\bibitem{barrow1} Barrow J. D. and Ottewill A. C., J. Phys. A: Math. Gen.
\textbf{16}, (1983) 2757.

\bibitem{barrow2} Barrow J. D. and Cotsakis S., Phys. Lett. \textbf{B214},
(1988) 515.

\bibitem{carroll} Carroll S. M. et al., Phys. Rev. \textbf{D70}, (2004)
043528.

\bibitem{starobinsky}Starobinsky A. A., JETP Lett. \textbf{86}, (2007) 157.

\bibitem{hu}Hu W. and Sawicki I., Phys. Rev. \textbf{D76}, (2007) 064004.

\bibitem{odintsov1}Nojiri S. and Odintsov S. D., arXiv:0807.0685.

\bibitem{odintsov2}Nojiri S and Odintsov S. D., Int. J. Geom. Meth.
Mod. Phys. \textbf{4}, (2007) 115.

\bibitem{capo} Capozziello S. and Francaviglia M., Gen. Relat. Gravit.
\textbf{40}, (2008) 357.

\bibitem{sotiriou} Sotiriou T. and Faraoni V., Rev. Mod. Phys. \textbf{82}, (2010) 451.

\bibitem{franco} Ferraro R. and Fiorini F., Phys. Rev. \textbf{D75}, (2007)
084031.

\bibitem{franco2} Ferraro R. and Fiorini F., Phys.Rev. \textbf{D78}, (2008)
124019.

\bibitem{tortion} Bengochea G. R. and Ferraro R., Phys. Rev. \textbf{D79}, (2009) 124019.

\bibitem{einstein} Einstein A., Sitzungsber. Preuss. Akad. Wiss. Phys. Math. KI.,(1928) 217;
ibid., (1930) 401; Einstein A., Math. Annal. \textbf{102}, (1930) 685.

\bibitem{hayashi} Hayashi K. and Shirafuji T., Phys. Rev. \textbf{D19},
(1979) 3524, Addendum-ibid. \textbf{D24}, (1982) 3312.

\bibitem{linder} Linder E. V., Phys. Rev. \textbf{D81}, (2010) 127301.

\bibitem{wu}Wu P. and Yu H., Phys. Lett. \textbf{B693}, (2010) 415.

\bibitem{wu2}Wu P. and Yu H., Phys. Lett. \textbf{B692}, (2010) 176.

\bibitem{wu3}Wu P. and Yu H., arXiv:1008.3669.

\bibitem{myr}Myrzakulov R., arXiv:1006.1120; arXiv:1008.4486.

\bibitem{yer}Yerzhanov K. K. et al., arXiv:1006.3879.

\bibitem{yang}Yang R. J., arXiv:1007.3571; arXiv:1010.1376.

\bibitem{karami}Karami K. and Abdolmaleki A., arXiv:1009.2459; arXiv:1009.3587.

\bibitem{dent}Dent J. B. et al., arXiv:1008.1250; arXiv:1010.2215.

\bibitem{lib}Li B. et al., arXiv:1010.1041.

\bibitem{bamba}Bamba K. et al., arXiv:1008.4036; Bamba K. et al., arXiv:1011.0508.

\bibitem{nesseris}Nesseris S. and Perivolaropoulos L., JCAP \textbf{0702}, (2007) 025.

\bibitem{wei1}Wei H., Phys. Lett. \textbf{B687}, (2010) 286.

\bibitem{li}Li M. et al., arXiv:0910.0717.

\bibitem{sollerman}Sollerman J. et al., Astrophys. J. \textbf{703}, (2009) 1374.

\bibitem{xu2} Xu L. and Wang Y., Phys. Rev. \textbf{D82}, (2010)
043503; Xu L., arXiv:1005.5055.

\bibitem{ohd39}Gaztanaga E. et al., MNRAS \textbf{399}, (2009) 1663.

\bibitem{ohd40}Riess A. G. et al., Astrophys. J. \textbf{699}, (2009) 539.

\bibitem{ohd41}Stern D. et al., JCAP \textbf{02}, (2010) 008.

\bibitem{weit} Weitzenb\"{o}ck R., \emph{Invarianten Theorie}, (Nordhoff,
Groningen, 1923).

\bibitem{maluf} Maluf J. W., J. Math. Phys. \textbf{35}, (1994) 335.

\bibitem{arcos} Arcos H. and Pereira J., Int. J. Mod. Phys. D\textbf{13},
(2004) 2193.

\bibitem{akaike} Akaike H., IEEE Trans. Auto. Control, \textbf{19}, (1974) 716.

\bibitem{liddle} Liddle A. R., MNRAS \textbf{351}, (2004) L49.

\bibitem{bies} Biesiada M., JCAP \textbf{702}, (2007) 003.

\bibitem{cal02}Caldwell R. R., Phys. Lett. \textbf{B545}, (2002) 23.

\bibitem{kessler}Kessler R. et al., Astrophys. J. Suppl. Ser. \textbf{185}, (2009) 32.

\bibitem{mlcs}Phillips M. M. et al., Astrophys. J. \textbf{413}, (1993)
L105; Riess A. G. et al., Astrophys. J. \textbf{438}, (1995) L17;
Jha S. et al., Astrophys. J. \textbf{659}, (2007) 122.

\bibitem{gb2}Bengochea G. R., Phys. Lett. \textbf{B} (2010), doi:10.1016/j.physletb.2010.12.014, arXiv:1010.4014.

\bibitem{eisen} Eisenstein D. et al., Astrophys. J. \textbf{633}, (2005) 560.

\bibitem{komatsu}Komatsu E. et al., Astrophys. J. Suppl. \textbf{180}, (2009) 330.

\bibitem{salt2} Guy J. et al., Astron. and Astrophys. \textbf{466}, (2007) 11.

\bibitem{lensing1}Jarvis M. et al., Astrophys. J. \textbf{644}, (2006) 71.

\bibitem{lensing2} Schrabback T. et al., Astron. and Astrophys. \textbf{516}, (2010) A63.

\bibitem{ghi}Ghirlanda E. et al., New J. Phys. \textbf{8}, (2006) 123.

\bibitem{basi}Basilakos S. and Perivolaropoulos L., MNRAS \textbf{391}, (2008) 411.

\bibitem{wei2}Wei H., JCAP \textbf{08}, (2010) 020.

\bibitem{xu1}Xu L. and Wang Y., arXiv:1007.4734.

\bibitem{chi2sn}Nesseris S. and Perivolaropoulos L., Phys. Rev. \textbf{D72}, (2005) 123519; Perivolaropoulos
L., Phys. Rev. \textbf{D71}, (2005) 063503.

\bibitem{paramR}Bond J. R. et al., MNRAS \textbf{291}, (1997) L33.

\bibitem{percival}Percival W. J. et al., MNRAS \textbf{401}, (2010) 2148.

\bibitem{schaefer} Schaefer B. E., Astrophys. J. \textbf{660}, (2007) 16.

\bibitem{amati1} Amati L. et al., Astron. and Astrophys. \textbf{390},
(2002) 81.

\bibitem{amati2}Amati L. et al., MNRAS \textbf{391}, (2008) 577.

\bibitem{amati3}Amati L., MNRAS \textbf{372}, (2006) 233.

\bibitem{amati4} Amati L. et al., Astron. and Astrophys. \textbf{508}, (2009) 173.

\bibitem{wang} Wang Y., Phys. Rev. \textbf{D78}, (2008) 123532.

\bibitem{ohd65}Jimenez R. and Loeb A., Astrophys. J. \textbf{573}, (2002) 37.

\bibitem{ohd66}Jimenez R. et al., Astrophys. J. \textbf{593}, (2003) 622.

\bibitem{ohd67} Simon J. et al., Phys. Rev. \textbf{D71}, (2005) 123001.


\end{thebibliography}
\end{document}